\documentclass[12pt,preprint]{aastex}

\shorttitle{Halo Central Mass and Phase-Space Densities}
\shortauthors{Shapiro \& Iliev}
\begin{document}
\title{The Central Mass
 and Phase-Space Densities of Dark Matter Halos: Cosmological
Implications}

\author{Paul R. Shapiro}
\affil{Department of Astronomy, University of Texas, Austin, 78712}
\email{shapiro@astro.as.utexas.edu}
\and
\author{Ilian T. Iliev}
\affil{
Osservatorio Astrofisico di Arcetri, Largo Enrico Fermi 5, 
50125 Firenze, Italy} 
\email{iliev@arcetri.astro.it}

\begin{abstract}
Current data suggest that the central mass densities $\rho_0$ and phase-space
densities $Q\equiv\rho_0/\sigma_V^3$ of cosmological halos in the present 
universe are correlated with their velocity dispersions $\sigma_V$ 
over a very wide range of $\sigma_V$ from less
than 10 to more than 1000 $\rm km\,s^{-1}$. Such correlations are an expected 
consequence of the statistical correlation of the formation epochs
of virialized objects in the CDM model with their masses; the smaller-mass
halos typically form first and merge to form larger-mass halos later.
We have derived the $Q-\sigma_V$ and $\rho_0-\sigma_V$ correlations for 
different CDM cosmologies and compared the predicted correlations with the 
observed properties of a sample of low-redshift halos ranging in
size from dwarf 
spheroidal galaxies to galaxy clusters. Our predictions are generally 
consistent with the data, with preference for the currently-favored, flat
$\Lambda$CDM model. Such a comparison serves to test the basic CDM paradigm
while constraining the background cosmology and the power-spectrum
of primordial density fluctuations, 
including larger wavenumbers than have previously been constrained. 
\end{abstract}


\section{Introduction}
Galaxies and clusters emerge in the CDM model when 
primordial density fluctuations grow to form dark-matter 
dominated halos in approximate virial equilibrium.
For Gaussian-random-noise density fluctuations in
a cosmologically-expanding, pressure-free gas of collisionless matter
like standard CDM, gravitational instability assembles these halos
hierarchically. Smaller-mass objects form first, then merge to form 
larger-mass objects, in a continuous sequence of increasing mass. 
As a result, a statistical correlation between the mass and collapse time 
of the objects which form is expected which depends upon
the shape and amplitude of the primordial power spectrum $P(k)=|\delta_k|^2$
and on the rates of fluctuation growth in different background cosmologies.
This suggests that a measurement of this correlation can constrain $P(k)$ and
the fundamental cosmological parameters, while testing the CDM model.

The power spectrum is already well constrained at large 
scales by observations of the anisotropy of the 
Cosmic Microwave Background (CMB) and the mass function
of galaxy clusters \citep{BOPS}. At large wavenumber $k$, however, 
$P(k)$ is still poorly constrained. Recent CMB anisotropy measurements
suggest that the universe is flat (e.g. de Bernardis et al. 2000, Hanany et 
al. 2000, Pryke et al. 2001). The measured 
distance-redshift relation for type Ia SNe (Perlmutter et al. 1999,
Riess et al. 2000; Reiss et al. 2001; Turner and Riess 2001) combines 
with this to favor a low-density universe with a cosmological constant
or some other ``dark energy''. It is important to establish whether
the statistical correlation of halo mass and formation epoch predicted 
at large $k$ by the standard CDM model is consistent with these 
constraints on $P(k)$ at small $k$ and on the fundamental cosmological 
parameters. 

Despite its many successes, the standard 
CDM model presently faces significant problems (e.g. Moore 2001 and 
refs. therein). For example, CDM simulations find singular halo
density profiles and high abundance of subhalos within halos, in 
apparent conflict with observations, and have difficulty explaining 
galactic rotation. Recently, modifications of the 
standard CDM model have been proposed 
to resolve some of these problems, including the suggestion of
self-interacting dark matter (SIDM) \citep{SS} and other CDM variants 
(e.g. see Wandelt et al. 2000 and refs. therein). In addition to testing 
the standard CDM paradigm while further constraining $P(k)$ at large
wavenumber, therefore, a measurement of the statistical correlations among
the internal properties of halos over a wide mass range from dwarf 
spheroidal (dSph) galaxies ($M_{\rm halo}\sim10^6-10^7M_\odot$) to galaxy 
clusters ($M_{\rm halo}\sim10^{14}-10^{15}M_\odot$), expected if halo masses
correlate with their formation epochs, would sharpen this debate.

For example, it was recently pointed out (Sellwood 2000, 
Dalcanton \& Hogan 2000, Hogan and Dalcanton 2000, Madsen 2001)
that the correlation of the 1D velocity dispersion $\sigma_V$ of a halo 
with the maximum value of its phase space density 
$f_{\rm max}\sim Q\equiv\rho_0/\sigma_V^3$, where $\rho_0$ is the central density of 
the halo, is a potent diagnostic of the properties of the dark matter. 
As argued in Sellwood (2000), Liouville's theorem applied to collisionless 
dark matter guarantees that processes like baryonic dissipation
will not change $f_{\rm max}$ significantly. In that case, $f_{\rm max}$
can be used as a diagnostic even for halos that are no longer
dark-matter dominated in the central region following
baryon cooling and compression. 
Of related interest, there are recent claims that the core densities of all 
galaxy and cluster halos are approximately the same, independent of the 
mass of the halo (Firmani et al. 2000, Avila-Reese et al. 2000, 
Kaplinghat, Knox, \& Turner 2000). A central density which is the same for
halos of different mass would impose strong constraints on the nature of 
the dark matter, if this hypothesis is correct.  

In what follows, we predict correlations of the 
maximum phase-space densities $Q$ and of the central densities $\rho_0$ of  
cosmological halos with their velocity dispersions $\sigma_V$, for different
CDM models. We will show that these predicted correlations are consistent with 
the observed properties of halos 
spanning a very large range of mass scales from dSph galaxies to galaxy 
clusters, and that the current data has the power to constrain the background 
universe and power spectrum of the CDM model. Finally, we shall discuss the
hypothesis that halos have a core density that is independent of their mass.

\section{The $Q-\sigma_V$ and $\rho_0-\sigma_V$ Correlations
for Cosmological Halos Observed Today}
We have developed an analytical model for the postcollapse equilibrium
structure of virialized objects which condense out of a 
cosmological background universe, either matter-dominated or flat with
a cosmological constant (Shapiro, Iliev \& Raga 1999, ``Paper I'';
Iliev \& Shapiro 2001b, ``Paper II''). The model is based upon the assumption
that cosmological halos form from the collapse and virialization of
``top-hat'' density perturbations and are spherical, isotropic, and 
isothermal. This leads to a unique, nonsingular, truncated
isothermal sphere (TIS), a particular solution of the Lane-Emden equation
(suitably modified when $\Lambda\neq0$). The size $r_t$ and velocity 
dispersion $\sigma_V$
are unique functions of the mass and redshift of formation of the object
for a given background universe. Our TIS density profile flattens to a 
constant central value, $\rho_0$, which is 
roughly proportional to the critical density of the universe at the 
epoch of collapse, with a small core radius $r_0\approx r_t/30$ (where 
$\sigma_V^2=4\pi G\rho_0r_0^2$ and $r_0\equiv r_{\rm King}/3$, for the ``King
radius'' $r_{\rm King}$, defined by Binney and Tremaine, 1987, p. 228).

This TIS model reproduces many of the average properties of the halos in
CDM N-body simulations quite well, suggesting that it is a useful approximation
for the halos which result from more realistic initial conditions. We have 
elsewhere compared this model with simulation results and applied it 
successfully in a variety of ways, including the derivation of the 
mass-radius-temperature virial relations and integrated mass profiles 
of X-ray clusters deduced from numerical CDM simulations by \citet{EMN} and
\citet{ME} (Papers I, II), the rotation curves of dark-matter dominated 
dwarf galaxies and the observed correlation between their maximum velocity,
${\rm v}_{\rm max}$, and the radius, $r_{\rm max}$, at which it occurs \citep{ISa},
and the mass model of cluster CL 0024 determined by
strong gravitational lensing 
measurements \citep{SI}. 
The TIS mass profile agrees well with the fit to N-body simulations by 
Navarro, Frenk \& White (1996; ``NFW'') (i.e. fractional deviation of 
$\sim 20\%$ or less) 
at all radii outside of a few TIS core radii (i.e. outside a King radius  
or so), for NFW concentration parameters $4\leq c_{\rm NFW}\leq7$. 
As a result, the TIS central density $\rho_0$ can be used to characterize the  
core density of cosmological halos, even if the latter have singular profiles
like that of NFW, as long as we interpret $\rho_0$, in that case,
as an average over the innermost region.

The TIS model yields the central density $\rho_0$, velocity dispersion 
$\sigma_V$, size $r_t$ and core radius $r_0$ for any 
given halo as unique functions of its mass $M$ and collapse redshift 
$z_{\rm coll}$. This defines a ``cosmic virial plane'' in the three-dimensional 
space of halo parameters ($\rho_0,r_0,\sigma_V$) [i.e. 
$2\ln\sigma_V-\ln\rho_0-2\ln r_0-\ln(4\pi G)=0$], in the terminology 
of Burstein et al. (1997), and determines the halo size, mass and collapse 
redshift for each point on this infinite plane, the ensemble of all 
possible virialized objects. In hierarchical models of structure formation
like CDM, however, the statistical correlation of halo mass and collapse
redshift determines the expected distribution of points on this cosmic 
virial plane. We shall combine our TIS model with the  
well-known Press-Schechter (PS) approximation \citep{PS} for 
$z_{\rm coll}(M)$ -- the typical collapse epoch for
a halo of mass M -- to derive the statistical distribution for the
$Q-\sigma_V$ and $\rho_0-\sigma_V$ correlations, as follows.

According to PS, the fraction 
of matter in the universe which is condensed into objects of mass 
$\geq M$ at a given epoch is
$
 	{\rm erfc}\left(\nu/2^{1/2}\right)$,
where $\nu\equiv{\delta_{\rm crit}}/\sigma(M)$, 
$\sigma(M)$ is the standard deviation of the density fluctuations  
at that epoch, according to linear theory, when filtered on mass
scale $M$, and  
$\delta_{\rm crit}$ is the fractional overdensity of 
a top-hat perturbation when this linear theory is extrapolated to the 
time of infinite collapse in the exact nonlinear solution. The ``typical''
collapse epoch for a given mass is that for which $\sigma(M)=\delta_{\rm crit}$
(i.e. $\nu=1$). For a given $z_{\rm coll}$, this defines a typical mass scale:
$M_\star\equiv M(\nu=1)$.

For any $\nu$, the PS prescription above yields a unique 
$z_{\rm coll}$ for halos of a given mass, thus completely determining the 
$Q-\sigma_V$ and $\rho_0-\sigma_V$ correlations.  The TIS+PS
predictions for the $Q-\sigma_V$ correlation for halos observed today
are shown in Figure~\ref{4panel_phase},  for $\nu=0.5-3$.
For any $\nu$, there is some value of $\sigma_V=\sigma_{V,0}$ for which $z_{\rm coll} = 0$.
For fluctuations which become halos observed today
with $\sigma_V \geq \sigma_{V,0}$, the most likely collapse epoch
is assumed to be $z_{\rm coll} = 0$. Hence,
curves of constant $\nu$ merge with the straight line for $z_{\rm coll} = 0$
for this range of $\sigma_V$ values.

For comparison, the observed properties of a sample of 
low-redshift, dark-matter dominated
halos of galaxies and clusters are also plotted in Figure~\ref{4panel_phase},
using data from the following sources. 
Kormendy and Freeman (1996; 2001) have compiled and 
interpreted mass models fitted to galaxy rotation curves for 49 
late-type spirals of type Sc-Im and velocity dispersions and core radii for 
7 dSph galaxies, chosen to minimize the need to correct the deduced
halo parameters
for the effects of baryonic dissipation. Data for one additional 
 dSph galaxy, Leo I, is from \citet{MOVK}. 
For low-redshift clusters, we have adopted central total mass densities 
which are based upon the central gas densities 
and baryonic-gas-mass fractions derived from
X-ray brightness profile fits for 28 nearby clusters by \citet{MME}, for which 
velocity dispersions were tabulated by Girardi et al. (1998) (26 clusters)
and \citet{JF} (2 clusters). The galaxy and cluster halo profiles adopted by 
these authors in fitting data to derive their central densities are very similar 
to the TIS halo profile we have used to make our predictions, so these central densities
are suitable for direct comparison.

The agreement in Figure~\ref{4panel_phase} between  our 
TIS+PS model predictions and the observed halos
is excellent for the entire range of halo masses  
from dSph galaxies to galaxy clusters. The observed scatter of the data is 
natural from a theoretical point of view, due to the Gaussian random 
statistics of the initial density fluctuations.
A comparison of results for different models suggests that
the currently-favored, flat $\Lambda$CDM model
($\Omega_0=0.3,\lambda_0=0.7, h=0.65$)
is preferred. The COBE-normalized, low-density, matter-dominated OCDM model 
($\Omega_0=0.3, \lambda_0=0, h=0.65$), with a primordial power 
spectrum index of $n_{\rm p}=1.3$, 
a tilt which allows it to match the galaxy cluster 
abundance at $z=0$ (e.g. Wang et al. 2000),
requires that most galaxy halos observed today
be low-$\nu$ (i.e. late-collapsing), rather than typical.
This may indicate that $P(k)$ for this model is too large on galactic mass
scales. 
The untilted OCDM
model ($n_{\rm p}=1$), which predicts an incorrect cluster
abundance at $z=0$ (i.e. not ``cluster-normalized''), requires, instead,
that all halos observed today
be high-$\nu$,
corresponding to rare, high-density peaks of the primordial 
density field.
In that case, $P(k)$ may be too small on cluster mass scales and below.
Finally, the cluster-normalized Einstein-de~Sitter (EdS)
model (``SCDM'') shows  good agreement with this halo data on galaxy scales, 
but there are many cluster data points for which the $Q$-values
fall below the limiting values predicted to correspond to $z_{\rm coll}=0$.

In Figure~\ref{4panel_rho0}, we compare the TIS+PS prediction
for the  $\rho_0 - \sigma_V$ correlation with the same halo data.
Once again, the data are generally consistent with the 
theoretical prediction, with the same trends for different CDM models 
as identified above for the $Q-\sigma_V$ correlation. Hence, the
flat, $\Lambda$CDM 
model is preferred. 
The data are clearly inconsistent, however, with the 
constant-core-density hypothesis, since observed core-densities show a 
clear downward trend from dSph galaxies to larger galaxies and clusters, 
decreasing by more than three orders of magnitude as $\sigma_V$ increases by
two orders of magnitude.
The appearance of a core-density that is approximately independent
of $\sigma_V$ at large values of $\sigma_V$ (i.e. from massive galaxy to 
cluster scales) is easily understood in the 
TIS+PS model if the Gaussian statistics of fluctuation amplitudes
are considered. Since large galaxies
and galaxy clusters observed today formed 
from high-$\nu$ fluctuations, the most probable 
time for them to collapse is close to the present. 
Therefore, our model predicts 
them all to have similar central densities, since $\rho_0$ is proportional
to the mean density of the universe at the collapse epoch. 

This success of the TIS+PS
model in predicting the observed correlations can be understood 
by an analytical argument, as follows. According to Paper II, the exact 
dependences of the TIS model core densities and velocity dispersions 
on $M$ and
$z_{\rm coll}$ are well approximated by 
$\rho_0=1.80\times10^4[F(z_{\rm coll}]^3\rho_{\rm crit,0}$ and 
$\sigma_V^2=1.1\times10^4(M/10^{12}M_\odot h^{-1})^{2/3}F(z_{\rm coll})
	{\rm\,km^2\,s^{-2}}$, where $\rho_{\rm crit,0}\equiv3H_0^2/(8\pi G)$
and the
function $F(z_{\rm coll})$ is defined by equation (84) of Paper II. 
The quantity $F^3$ is just the density of a top-hat perturbation 
after it collapses and virializes in the standard uniform sphere
(SUS) approximation (as measured in units of the EdS value for
$z_{\rm coll}=0$, 
$18\pi^2\rho_{\rm crit,0}$).  
For the EdS case, $F=(1+z_{\rm coll})$, 
while for the low-density, matter-dominated and flat cases, 
$F\rightarrow\Omega_0^{1/3}(1+z_{\rm coll})$ for early collapse.
If we approximate the power-spectrum of density fluctuations at high
redshift (e.g. just after recombination) as a power-law in wavenumber $k$,
$P(k)\propto k^{n}$, 
and define a mass $M\propto k^{-3}$, then
 $F\propto M^{-(3+n)/6}$ if 
$n=n_{\rm eff}\equiv-3(2y_F+1)$, where 
$y_F\equiv(d\ln F/d\ln M)_{\rm exact}$ at the relevant
mass scale. For all masses in the EdS case and for masses which collapse
early in the low-density,   matter-dominated and flat, universes, $y_F$ reduces 
to $y_\sigma\equiv(d\ln \sigma/d\ln M)_{\rm exact}$, and  
$(1+z_{\rm coll})\propto\sigma(M)\propto M^{-(3+n)/6}$, where $\sigma(M)$ is
evaluated at the same cosmic time for all masses.
The dependence of $n_{\rm eff}$ and $z_{\rm coll}$ on $M$
for $1$-$\sigma$ fluctuations is shown in Figure~\ref{n_eff_lam} for $\Lambda$CDM,
along with 
the approximate $n_{\rm eff}$ which results if $y_F$ is replaced by $y_\sigma$, 
which shows that the latter is a very good approximation for all masses 
$M<10^{12}M_\odot h^{-1}$. In terms of this power-law model, the TIS model 
then yields $\rho_0\propto M^{-(n+3)/2}$, $\sigma_V\propto M^{(1-n)/12}$,
and $Q\propto M^{-(n+7)/4}$, which combine to give
\begin{equation}
\label{scaling_Q}
Q=Q_{\star}(\sigma_V/\sigma_{V,\star})^{\alpha-3},
\end{equation} 
and
\begin{equation}
\label{scaling_rho0}
\rho_0=\rho_{0,\star}(\sigma_V/\sigma_{V,\star})^{\alpha},
\end{equation}
where $\alpha\equiv 6(n+3)/(n-1)$,  
except that $\alpha=0$ for $M>M_\star(z=0)$, 
for which $z_{\rm coll}=0$ is assumed.

For the currently-favored, flat, $\Lambda$CDM model,
for example, $n_{\rm eff}\approx -2.6, -2.4, -2.2$ for 
$M=M_\star=10^7, 10^{10}, 10^{12}M_\odot h^{-1}$, respectively, and, hence,
$\alpha\approx-0.7, -1.1, -1.8$ for these masses, while 
$\sigma_{V,\star}=4.5, 33, 114\, {\rm km\,s^{-1}}$, 
$Q_\star= 1.4\times10^{-3}, 5.7\times10^{-7}, 2.4\times10^{-9}
	\,M_\odot \mbox{pc}^{-3}(\mbox{km}\,\mbox{s}^{-1})^{-3}$, and
$\rho_{0,\star}= 0.12, 0.02, 0.0035 \,M_\odot\mbox{pc}^{-3}$, respectively.
For all masses $M > M_*(z=0) \approx 10^{13} h^{-1} M_\odot$, our model
predicts $\rho_0(z_{\rm coll}=0)=0.00143\,M_\odot\mbox{pc}^{-3}$, and 
$Q(z_{\rm coll}=0)=\rho_0(z_{\rm coll}=0)/\sigma_V^{3}
$.

Dalcanton and Hogan (2000) predict that 
$Q\propto\sigma_V^{-3}$ for all $\sigma_V$,
 which in our model is achieved only for $M\rightarrow0$,
for which $n_{\rm eff}=-3$,  and in the high-mass limit, for which
$z_{\rm coll}\approx 0$ typically. Similarly,
according to equation (\ref{scaling_rho0}), the TIS + PS model predicts 
that the core-density is the same for halos with different $\sigma_V$ only in
the small-mass limit in which $n_{\rm eff}=-3$ and  
the high-mass limit, for which $z_{\rm coll}\approx 0$,
with a large difference between the values
of $\rho_0$ in these two limits.  The generally good agreement
between this prediction and the data, as plotted in Figure 2, argues against
the hypothesis that all halos share a single, universal core density. 
The putative universal core density may have escaped detection 
by the observations quoted here if it is confined to a region interior
to the smallest radii probed by those observations, but it must be at least 
as large as the highest central densities reported in Figure~\ref{4panel_rho0}.
These high central densities are well in excess, of the universal 
core density of $0.02 M_\odot {\rm pc}^{-3}$ claimed by \citet{FDCHA}.

In the future, as more and better data become available, it should be
possible to refine the comparison between the correlations predicted here
and the observed properties of galaxies and clusters.  We have
demonstrated that such a correlation is expected, 
that current data are 
consistent with the predicted correlations for the CDM model, with preference 
for the currently-favored $\Lambda$CDM model 
over CDM in other background cosmologies, and that
no significant deviation of the primordial power-spectrum shape from the
 scale-invariant, Harrison-Zeldovich shape ($n_p=1$), the
standard  prediction of inflationary cosmology, is required.
In addition, we have demonstrated that the data do not
support the idea that all halos have the same core density,
independent of their mass or formation epoch.


We thank Hugo Martel, Sandra Faber
and Ken Freeman for discussion and John Kormendy for providing his tabulation
of data points for galactic halos used here. 
This research was supported by grants NASA ATP NAG5-7363, 
-7821, and -10825, TARP 3658-0624-1999, 
European Community RTN contract HPRN-CT2000-00126 RG29185, and
NSF INT-0003682.

\figcaption[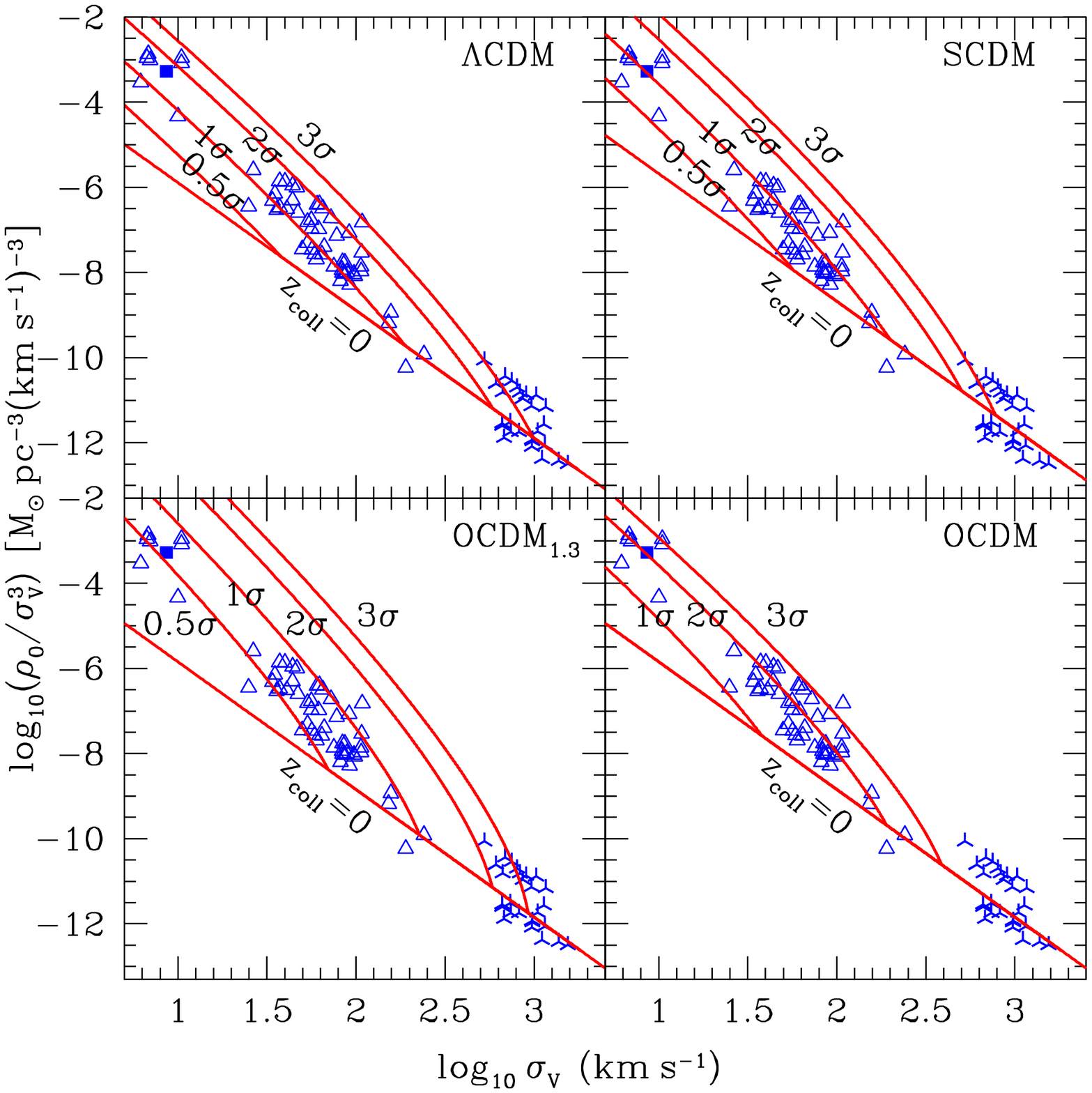]
{Maximum phase-space density $Q=\rho_0/\sigma_V^3$
versus velocity dispersion $\sigma_V$ for halos observed today,
as predicted for various CDM universes by the
TIS + PS model (solid curves) for halos formed from
$\nu$-$\sigma$ fluctuations, as labelled with the values of $\nu$, for 
$\nu =$
0.5, 1, 2, and 3.  In each panel, lines representing halos of different
mass which collapse at the same redshift are shown for the case 
$z_{\rm coll} = 0$, 
as labelled.  Each panel represents different assumptions for the 
background universe and primordial density fluctuations, 
as labelled: COBE-normalized
$\Lambda$CDM ($\lambda_0=0.7$, $\Omega_0=0.3$) (upper left), 
cluster-normalized SCDM
($\Omega_0=1$) (upper right), and COBE-normalized OCDM ($\Omega_0=0.3$)
(lower panels), all assuming
$h=0.65$ and primordial power spectrum index $n_p =$1 (i.e. untilted),
except for OCDM$_{1.3}$, for which $n_p =$1.3.
Data points represent observed
galaxies and clusters, taken from the following sources: 
(1) galaxies from Kormendy \& Freeman (1996; 2001)
(open triangles); 
(2) Local Group dwarf galaxy Leo I from Mateo et al. (1998) (filled square);
(3) cluster velocity dispersions from Girardi et al. (1998) and Jones 
\& Forman (1999), central densities from Mohr et al. (1999) (crosses).
\label{4panel_phase}}

\figcaption[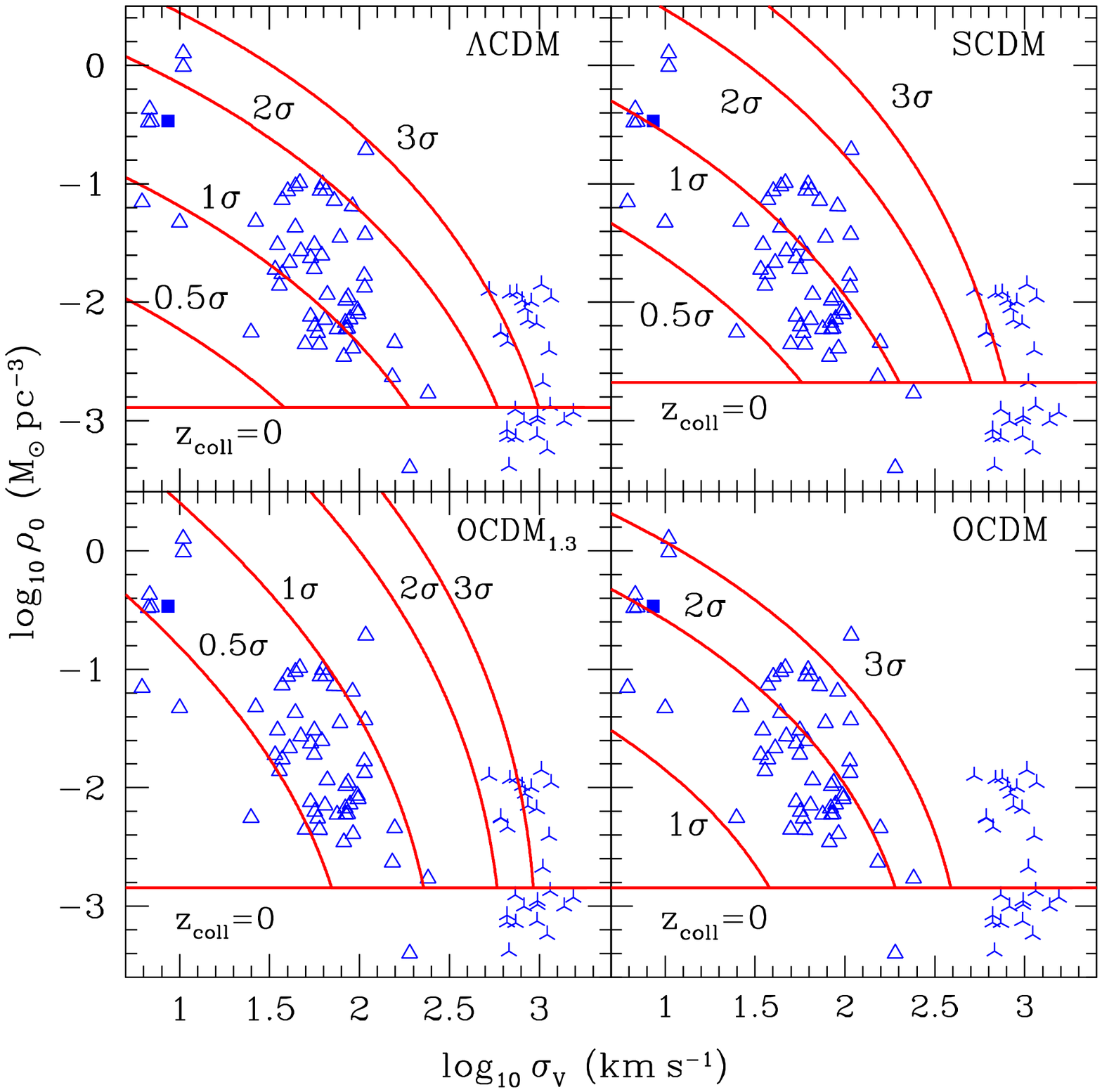]{Same as Figure~\ref{4panel_phase}, but for 
central density $\rho_0$ versus velocity dispersion $\sigma_V$.
\label{4panel_rho0}}

\figcaption[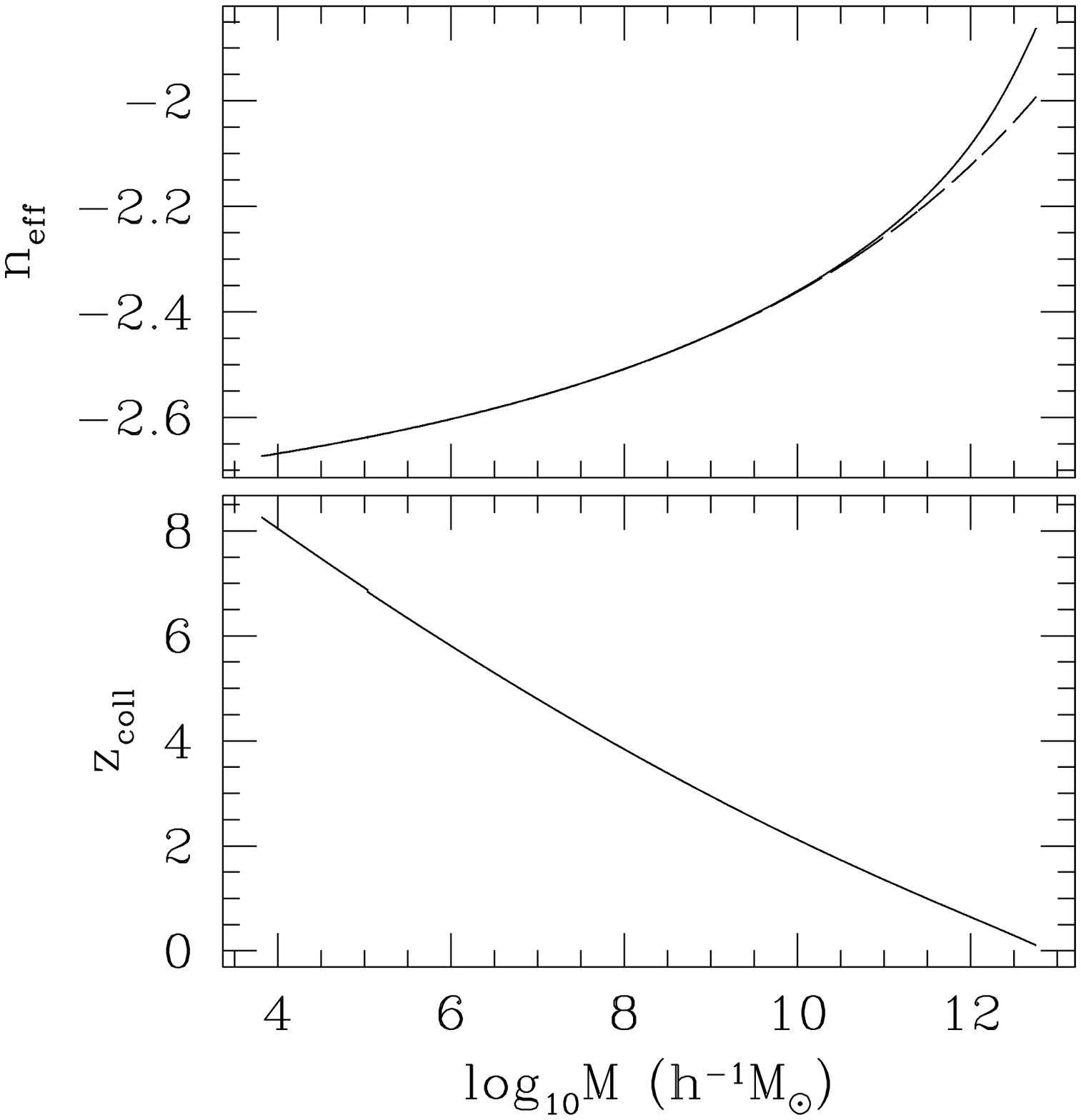]{ (a)(upper)
	Effective logarithmic slope, $n_{\rm eff}$, of the density 
	fluctuation power spectrum versus halo mass $M=M_\star$ (i.e. for
	$1$-$\sigma$ fluctuations) for $\Lambda$CDM, based upon $y_F$ (solid curve)
	and the approximation which uses $y_\sigma$, instead (dashed).    
	(b) (lower) Typical collapse redshift $z_{\rm coll}$ 
        for halos of mass $M_\star$ in $\Lambda$CDM.
\label{n_eff_lam}}

\begin{figure}
\epsscale{0.6}
\plotone{f1.eps}
\end{figure}

\begin{figure}
\epsscale{0.6}
\plotone{f2.eps}
\end{figure}

\begin{figure}
\epsscale{0.5}
\plotone{f3.eps}
\end{figure}

\end{document}